\documentclass[aps,pra,twocolumn,groupedaddress,showpacs]{revtex4}

\usepackage{graphicx}
\usepackage{graphicx}
\usepackage{amssymb}
\usepackage{amsmath}
\usepackage{color}
\newcommand{\be}{\begin{equation}}
\newcommand{\ee}{\end{equation}}
 
 \definecolor{BrickRed}{cmyk}{0,0.89,0.94,0.28}
\definecolor{MidnightBlue}{cmyk}{0.98,0.13,0,0.43}
\definecolor{DarkGreen}{rgb}{0,0.7,0.1}

\begin{document}

\title{Proposal to measure the Casimir Effect Across the Superconducting Transition}


\author{Giuseppe Bimonte}

\affiliation{ Dipartimento di Fisica E. Pancini, Universit\`{a} di
Napoli Federico II, Complesso Universitario
di Monte S. Angelo,  Via Cintia, I-80126 Napoli, Italy}
\affiliation{INFN Sezione di Napoli, I-80126 Napoli, Italy}

\email{giuseppe.bimonte@na.infn.it}

\begin{abstract}

The influence of the superconducting transition on the Casimir effect remains experimentally elusive. We propose a novel approach to address this challenge, exploiting the first-order superconducting transition induced in a thick film by a parallel magnetic field. Our calculations for a Au sphere opposite a Pb film,  based on the Mattis-Bardeen theory, predict a  jump in the Casimir force across the transition.  By periodically modulating the magnetic field, we induce a corresponding modulation of the Casimir force that should be readily detectable with existing micro-torsional oscillator technology, opening a promising pathway to finally observe the long-sought Casimir-superconductivity coupling. A successful measurement of this effect would provide crucial insights into the fundamental nature of the Casimir force and its interaction with macroscopic quantum phenomena.

\end{abstract}

\pacs{12.20.-m, 
03.70.+k, 
42.25.Fx 
}

\maketitle

\section{Introduction}
\label{sec:intro}

Superconductivity \cite{tinkham,buckel} and the Casimir effect \cite{book1,parse,book2,RMP,capasso,buh,woods,mehran} are striking examples of quantum phenomena observable on a macroscopic scale.  The investigation of their interplay has thus become a topic of considerable interest \cite{bimontesuper,inui,villareal,castillo2021,allocca2022,castillo}.  However, observing the influence of superconductivity on the Casimir force is experimentally challenging due to the nature of the Casimir force itself.  This force diminishes rapidly with the separation ($d$) between interacting bodies, limiting precise measurements to sub-micron distances.  At these separations, the relevant quantum and thermal fluctuations of the electromagnetic field have an energy scale ($\hbar c/ d$) in the eV range.  This is significantly larger than the characteristic energy scale of superconductivity ($k_B T_c$), which for a typical superconductor with a critical temperature of a few Kelvin lies in the far-infrared region.  As a result, the expected change in the Casimir force due to the superconducting transition is estimated to be less than $4.4 \times  10^{-4}(T_c /K)(d/\mu m$), a tiny fraction that pushes the limits of current experimental capabilities.

Several attempts have been made to observe the influence of the superconducting transition on the Casimir effect, thus far without conclusive results.  
Early work by Bimonte et al. \cite{ala1,ala2}  explored the Casimir-induced shift in the critical magnetic field ($H_ c$) of a thin superconducting film within a rigid Casimir apparatus.  An experiment using an Aluminum film placed an upper bound on this shift, consistent with theoretical predictions \cite{superc,annalisa}.  Later, an experiment by the Leiden group \cite{leiden} measured the Casimir force between a gold-coated sphere and a superconducting NbTiN film. While room-temperature data agreed well with theory, low-temperature measurements exhibited an unexplained 20 \% increase in force.  Crucially, no change in the Casimir force across the superconducting transition was observed, establishing an upper bound of 2.6\% on any such variation.
More recently, microfabricated platforms have been employed.  A study by Norte et al. \cite{norte} used two parallel, superconducting Aluminum-coated silicon nitride strings to create cavities with sub-100 nm separations.  By monitoring the resonance frequency of an optomechanical cavity, they aimed to detect changes in the Casimir force as the system transitioned to superconductivity.  The null result yielded an upper bound of 6 mPa on the pressure variation,  consistent with subsequent theoretical work \cite{bimonte2019}.  Bishop et al. \cite{bishop}  investigated the influence of the Casimir interaction between a gold and a lead film on the lead's critical temperature.  No shift in $T_c$ greater than 12 $\mu$K was observed down to ~70 nm spacing at zero magnetic field.
Finally, a recent experiment by Lepinay et al. \cite{lepin}  examined the nonlinear dynamics of a superconducting drum resonator in a microwave optomechanical system.  Using an Aluminum cavity with an 18 nm gap at 10 mK, they reported excellent agreement between experimental data and a power-law fit ($P_c= A/d^n$, with $A=(696 \pm 4) \times 10^5$ Pa and  $n=3.193 \pm 0.005$, with $d$  in nm) to the theoretical Casimir pressure, calculated using the BCS formula \cite{mattis}.   However, our independent analysis suggests that the fitted Casimir pressure in the superconducting state is statistically indistinguishable from that in the normal state, precluding any conclusions about the superconducting transition's effect on the Casimir force.  
 
 Motivated by the continued absence of experimental confirmation of the superconducting transition's influence on the Casimir effect, we propose a novel detection scheme. This Letter details this scheme, which leverages the fact that a  type-I superconducting film  (thickness $>>$  penetration depth  $\lambda$) below $T_c$ undergoes a first-order transition in a sufficiently strong parallel magnetic field $H_c$, rather than the usual second-order transition \cite{tinkham}.  Based on BCS theory \cite{mattis}, 
we predict a finite jump, $\Delta F_{\rm C}$ in the Casimir force across this transition, contrasting with the null variation in zero field. By modulating $H$ near $H_c$, we can induce a periodic variation of the Casimir force with amplitude $\Delta F_{\rm C}$. This modulation allows us to utilize the exceptional sensitivity of micro-torsional oscillators \cite{ricardomag, ricardotren}, which have demonstrated sub-fN sensitivity in Casimir force measurements.  Our calculations for a Pb film, presented below, demonstrate that readily achievable magnetic fields can generate a $\Delta F_{\rm C}$ large enough for reliable detection.

\begin{figure}
\includegraphics [width=.9\columnwidth]{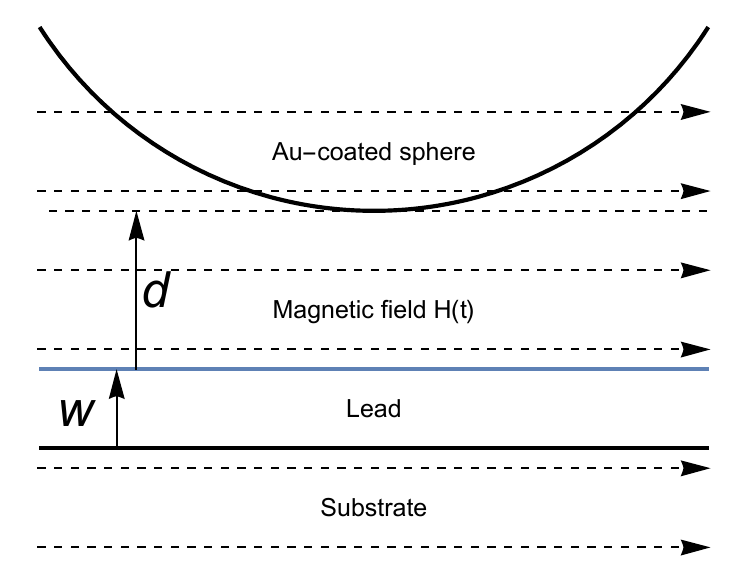}
\caption{\label{setup}  
Schematic of the Casimir setup: A gold-coated sphere (radius $R$) positioned a distance $d$ from a thick lead film (thickness $w >>$  London penetration depth  $\lambda$) deposited on a substrate. The Casimir force is modulated by a time-varying parallel magnetic field, $H(t)$, driving the lead film between its normal and superconducting states around the critical field $H_c(T)$.} 
\end{figure}

We consider the experimental Casimir configuration illustrated in Fig. \ref{setup}, comprising a gold-coated sphere with radius $R=150\,\mu$m positioned at a distance $d \ll R$ from a lead film of thickness $w$ supported by a substrate.   The entire system is immersed in a parallel time-varying magnetic field $H(t)$. This sphere radius is consistent with those used in recent micro-torsional oscillator Casimir experiments \cite{ricardomag, ricardotren}.

Before calculating the Casimir force, we review some essential facts about superconductors.
As is well known \cite{tinkham} at temperatures $T < T_c$ a type-I superconductor  transitions to the normal state in a parallel magnetic field $H$ exceeding the critical field $H_c(T)$. Crucially, for samples much thicker than the London penetration depth $\lambda(T)$, the transition is of first order,  contrasting with the second-order transition observed in zero field. $H_c(T)$ is empirically described by the parabolic equation:
\be
H_c(T) \approx H_c(0) \left[1- (T/T_c)^2 \right]\;, \label{HT}
\ee
where $H_c(0)$ represents the thermodynamic critical field.  The temperature dependence of $\lambda(T)$ is approximated by:
\be
\lambda(T)=\lambda(0) \left[1- (T/T_c)^4 \right]^{-1/2}\;.
\ee
For Pb, $T_c=7.2$ K, $H_c(0)=800$ Oe, $\lambda(0)= 32 \div 39$ nm \cite{buckel}.
We consider a fixed temperature $T$ below $T_c$ and induce periodic transitions between the normal and superconducting states by applying a time-dependent magnetic field $H(t)$ of the form \be
H(t) = H_c(T) + h \,f(t)\;,\label{Hmod}
\ee
where $f(t)$ represents a  unit amplitude rectangular pulse 
 with frequency $\nu$ \footnote{With ${\bar T}= 1/\nu$ denoting the pulse period, $f(t)$ is defined as a periodic function, $f(t + {\bar T}) = f(t)$, that is odd, $f(-t) = -f(t)$, and takes the value 1/2 for $0 < t < {\bar T}/2$.} and $h$ is a small magnetic field. Experiments have demonstrated sub-microsecond transition times in time-varying magnetic fields \cite{nether}.   Given that our chosen modulation frequencies  ($\nu$ ~ hundreds of Hz)  will match the resonance frequencies of the micro-torsional oscillators employed in recent Casimir experiments \cite{ricardomag, ricardotren}, we can safely approximate the superconducting transition as instantaneous across the edges of the rectangular magnetic field pulses.

From the foregoing considerations, it follows that the Casimir force $F_{\rm C}(t)$, in the setup of Fig. \ref{setup} will exhibit a time dependence given by:
\be
F_{\rm C}(t) =  \bar{F}_{\rm C}+   \Delta F_{\rm C}\,f(t) \;,
\ee
where $\bar{F}_{\rm C}= [F^{(\rm n)}_{\rm C}+F^{(\rm s)}_{\rm C}]/2$ and 
\be
\Delta F_{\rm C}= F^{(\rm n)}_{\rm C}-F^{(\rm s)}_{\rm C}
\ee 
represent respectively  the average and the  difference   of the Casimir forces in the normal and superconducting states at temperature $T$. Our signal of interest  is $ \Delta F_{\rm C}$.

 We now proceed to the calculation of $ \Delta F_{\rm C}$.  Given that the sphere radius R is much larger than the tip-plate separation $d$, the Proximity Force Approximation (PFA) provides a suitable method for estimating the difference in Casimir force.  The error incurred by the PFA in the sphere-plate geometry is less than $d/ R$  \cite{bimonte2012,ingold}. The PFA yields:
\be
\Delta F_{\rm C}= 2 \pi R \; \left[ {\cal F}_{\rm pp}^{(n)}(T,d)-{\cal F}_{\rm pp}^{(s)}(T,d) \right]\;.\label{PFA}
\ee
Here, ${\cal F}_{\rm pp}^{(n/s)}(T,d)$ represents the unit-area free energy for two parallel slabs of Au and Pb, respectively, separated by a distance $d$.    Our assumption that the Au coating and Pb film thicknesses are much greater than the skin depth justifies neglecting the influence of the underlying substrates.  The Lifshitz formula \cite{lifs} provides the expression for ${\cal F}_{\rm pp}^{(n|s)}(T,d)$:
$$
{\cal F}^{(n/s)}(T,d)=\frac{k_B T}{2 \pi}\sum_{l=0}^{\infty}\left(1-\frac{1}{2}\delta_{l0}\right)\int_0^{\infty} d k_{\perp} k_{\perp}  
$$
\be
\times \; \sum_{\alpha={\rm TE,TM}} \log \left[1- {e^{-2 d q_l}}{R_{{\rm Au}|\alpha}({\rm i} \xi_l,k_{\perp})\;R^{(n/s)}_{{\rm Pb}|\alpha}({\rm i} \xi_l,k_{\perp})} \right]\;,\label{lifs}
\ee
where $k_B$ denotes the Boltzmann constant, $\xi_l=2 \pi l k_B T/\hbar$ represent the (imaginary) Matsubara frequencies, and $k_{\perp}$ is the magnitude of the in-plane wave-vector. We define $q_l=\sqrt{\xi_l^2/c^2+k_{\perp}^2}$.  $R_{{\rm Au}|\alpha}({\rm i} \xi_l,k_{\perp})$  and $R^{(n/s)}_{{\rm Pb}|\alpha}({\rm i} \xi_l,k_{\perp})$ are the reflection coefficients of  Au or Pb slabs, respectively, for polarization $\alpha$. The superscript ${n/s}$ in $R^{(n/s)}_{{\rm Pb}|\alpha}$  indicates the normal or superconducting state of the Pb film.  The reflection coefficients are given by the Fresnel formulas:
\be
R_{\rm TE}(i \xi_l, k_{\perp}) =\frac{q_l-s_l}{q_l+s_l}\;,\label{TE}
\ee
\be
R_{\rm TM}(i \xi_l, k_{\perp}) =\frac{\epsilon_l q_l-s_l}{\epsilon_l\,q_l+s_l}\;,\label{TM}
\ee
where $s_l=\sqrt{\epsilon_l \xi_l^2/c^2+k_{\perp}^2}$, $\epsilon_l \equiv \epsilon(i \xi_l)$   is the material's electric permittivity along the imaginary frequency axis. 
A key point regarding the $ \Delta F_{\rm C}$ calculation is that, while precise computation of the individual Casimir free energies,   ${\cal F}^{(n)}(T,d)$ and ${\cal F}^{(s)}(T,d)$, requires Matsubara sums in Eq. (\ref{lifs}) extending to frequencies $\xi_l  > 10 \,c/d$, far fewer terms are needed for the difference,  ${\cal F}^{(n)}(T,d)-{\cal F}^{(s)}(T,d)$. This is because the reflection coefficients in the normal and superconducting states differ significantly only for frequencies below a few  $k_B T_c/\hbar$ \cite{tinkham}, which are in the far-infrared.  The far-infrared nature of these relevant Matsubara modes justifies neglecting core electron contributions to the Au and Pb permittivities and considering only intraband transitions, which are well described by a Drude dielectric function:
\be
\epsilon(i \xi)=1+ \frac{\Omega^2}{\xi (\xi + \gamma)}\;,\label{drude}
\ee   
\begin{figure}
\includegraphics [width=.9\columnwidth]{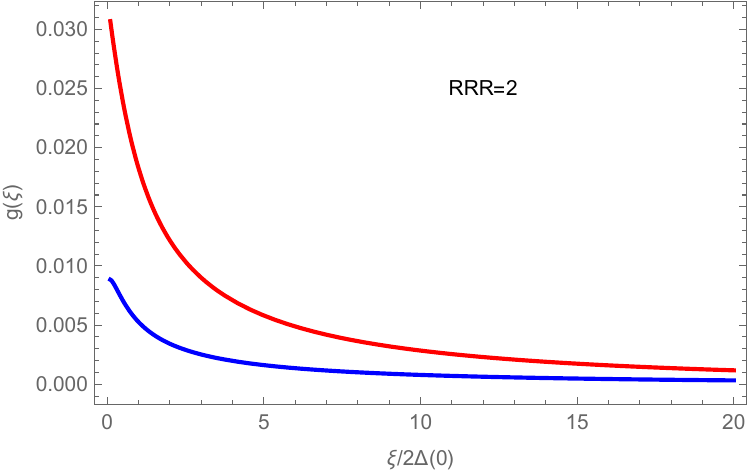}
\caption{\label{gfun}  Plot of the Mattis-Bardeen function $g(\xi)$  for Pb (RRR=2), as a function of $\xi/2 \Delta(0)$  at $T/T_{\rm c}=0.9$ (lower blue line) and $T/T_{\rm c}=0.1$ (upper red line).} 
\end{figure}
where  $\Omega$ represents the  plasma frequency for intraband transitions, and $\gamma$ is the relaxation frequency.   
In our calculations, we have assumed the plasma frequency, $\Omega$, to be temperature-independent, using its room-temperature value.  The relaxation frequency, $\gamma$, on the other hand, varies with temperature, typically decreasing as temperature decreases. At cryogenic temperatures, $\gamma$ approaches a constant residual value that depends on the specific sample.  Following standard practice, we relate the residual relaxation frequency to the room-temperature value, $\gamma_0$, using the formula $\gamma = \gamma_0/{\rm RRR}$, where RRR is the residual resistance ratio. The following Drude parameters were used: for Au  $\Omega =9$ eV/$\hbar$ and  $\gamma_0= 35$ meV/$\hbar$ \cite{book2}; for Pb, $\Omega=7.36$ eV/$\hbar$,  $\gamma_0= 200$ meV/$\hbar$  \cite{ordal}. 
\begin{figure}
\includegraphics [width=.9\columnwidth]{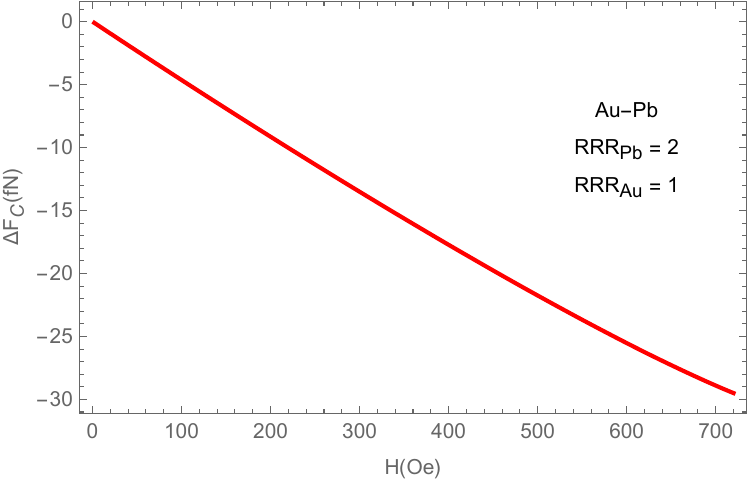}
\caption{\label{DeltaF}    Calculated change in Casimir force (fN) across the superconducting transition in an external magnetic field $H$ (Oe), for $R=150\, \mu$m and  $d=70$ nm. The transition temperature $T'_c$ is given by  $T'_c=T_c \sqrt{1-H/H_c(0)}$ (Eq. (\ref{HT}))} 
\end{figure}

The Drude model (Eq.(\ref{drude})) was employed for Au and Pb in its normal state.  The permittivity of Pb in the superconducting state was calculated using the Mattis-Bardeen formula for the conductivity $\sigma_{\rm BCS}(\omega)$ \cite{mattis}, known to accurately represent the optical response of BCS superconductors \cite{tinkham}.  While the general Mattis-Bardeen formula depends on both frequency ($\omega$) and wavevector ($q$), the $q$-dependence is negligible in the dirty limit ($\ell / \xi_0 \ll 1$), where  $\ell=v_F/\gamma$  is the mean free path and  $\xi_0=\hbar v_F/\pi \Delta(0)$ is the correlation length ($v_F$ being the Fermi velocity).  For Pb, $\ell / \xi_0 =0.07$, satisfying the dirty limit condition.  Analytic continuation to imaginary frequencies \cite{bimonteBCS}, gives:
\be
\sigma_{\rm BCS}(i \xi)=\frac{\Omega^2}{4 \pi}  \left[\frac{1}{(\xi+\gamma)}+  \frac{g(\xi;T)}{\xi}\right]\;,\label{MB}
\ee
The first term is the Drude contribution, and the second is the BCS correction. The explicit expression of $g(\xi)$ and its properties are discussed in  \cite{bimonte2019}.  
The permittivity of superconducting Pb is then given by:
\be
\epsilon_{\rm BCS}(i \xi)= 1+\frac{4 \pi}{\xi}\sigma_{\rm BCS}(i \xi)=1+ \frac{\Omega^2}{\xi }\left[\frac{1}{\xi+ \gamma}+   \frac{g(\xi;T)}{\xi}\right]\;.\label{BCS}
\ee
Fig. \ref{gfun} shows  the Mattis-Bardeen function $g(\xi)$ for Pb (RRR=2) vs. $\xi/2 \Delta(0)$  at $T/T_{\rm c}=0.9$ (lower blue line) and $T/T_{\rm c}=0.1$ (upper red line).  Here $2 \Delta(0)=3.528 \,k_B T_c= 2.2$ meV is the $T=0$ BCS gap. $g(\xi)$ becomes negligible for frequencies $\xi$ greater than a few tens of $2 \Delta (0)$, indicating that the superconducting permittivity becomes indistinguishable from the normal-state permittivity at these frequencies.

\begin{figure}
\includegraphics [width=.9\columnwidth]{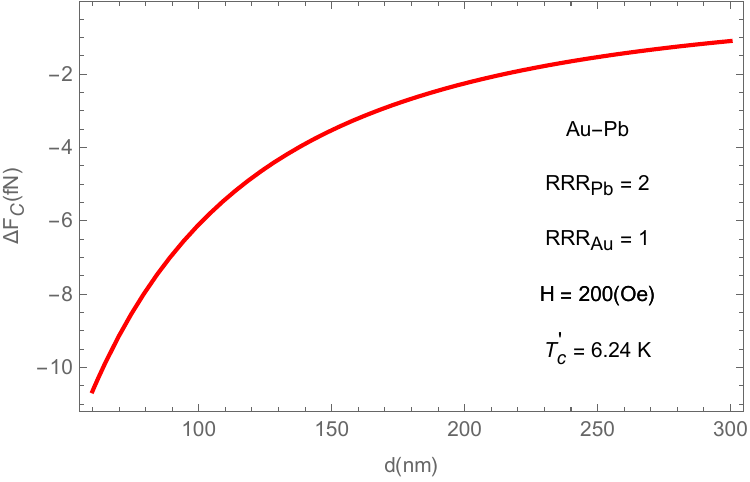}
\caption{\label{DeltaFvsd}    Casimir force change, $\Delta F_C$ (fN), across the superconducting transition vs. separation $d$ (nm) at $H = 200$ Oe ($T'_c=6.24$ K), calculated for $R=150\, \mu$m.} 
\end{figure}

Figure \ref{DeltaF} displays the calculated change in Casimir force, $\Delta F_C$(fN), across the superconducting transition   as a function of applied magnetic field $H$ (Oe) for $R=150\, \mu$m and   $d = 70$ nm. The relationship between the transition temperature, $T'_c$, and the magnetic field $H$ is given by  $T'_c=T_c \sqrt{1-H/H_c(0)}$ (Eq. (\ref{HT})).  The computation assumes residual resistance ratios of RRR = 2 for the Pb film and RRR = 1 for the Au coating.  Figure \ref{DeltaFvsd} displays $\Delta F_C$ (fN), vs. separation $d$ (nm) at $H = 200$ Oe ($T'_c=6.24$ K).  With the sub-fN sensitivity of micro-torsional oscillators used in recent Casimir experiments  \cite{ricardomag, ricardotren}, these figures demonstrate that $\Delta F_C$  is readily measurable across the transition, even with moderate magnetic fields and over a wide range of separations.

We have proposed and analyzed a novel approach to experimentally detect the elusive influence of the superconducting transition on the Casimir force. By exploiting the first-order superconducting transition induced by a modulated magnetic field in a thick film, we predict a measurable modulation of the Casimir force. Our calculations, using the Mattis-Bardeen theory for the superconducting state, demonstrate the feasibility of detecting this modulated signal with existing micro-torsional oscillator technology. This approach offers a promising pathway to finally observe the long-sought Casimir-superconductivity coupling.

\end{document}